\documentclass[iop]{emulateapj}
\usepackage{apjfonts}

\newcommand{\hi}{H\,{\sc i}}
\newcommand{\dol}{\textsc{dolphot}}
\newcommand{\n}{NGC }
\newcommand{\msol}{$M_{\sun}$}

\shorttitle{The \n4565 Disk Warp}
\shortauthors{Radburn-Smith et al.}
\submitted{Received 2013 July 30; accepted 2013 November 13}
\journalinfo{The Astrophysical Journal}

\begin{document}

\title{Constraining the age of the \n4565 \hi{} Disk Warp: Determining the
  Origin of Gas Warps}

\author{David~J.~Radburn-Smith\altaffilmark{1},
  Roelof~S.~de~Jong\altaffilmark{2},
  David~Streich\altaffilmark{2},
  Eric~F.~Bell\altaffilmark{3},
  Julianne~J.~Dalcanton\altaffilmark{1},
  Andrew~E.~Dolphin\altaffilmark{4},
  Adrienne~M.~Stilp\altaffilmark{1},
  Antonela~Monachesi\altaffilmark{3},
  Benne~W.~Holwerda\altaffilmark{5},
  and Jeremy~Bailin\altaffilmark{6}}

\affil{$^1$Department of Astronomy, University of Washington, Seattle, WA 98195, USA}
\affil{$^2$Leibniz-Institut f\"ur Astrophysik Potsdam, D-14482
  Potsdam, Germany}
\affil{$^3$Department of Astronomy, University of Michigan, Ann Arbor, MI 48109, USA}
\affil{$^4$Raytheon, 1151 East Hermans Road, Tucson, AZ 85756,
  USA}
\affil{$^5$European Space Agency, ESTEC, 2200 AG Noordwijk, The Netherlands}
\affil{$^6$Department of Physics and Astronomy, University of Alabama, Tuscaloosa, AL 35487, USA}

\begin{abstract}
  We have mapped the distribution of young and old stars in the
  gaseous \hi{} warp of \n4565. We find a clear correlation of young
  stars ($<600$ Myr) with the warp, but no coincident old stars
  ($>1$ Gyr), which places an upper limit on the age of the structure.
  The formation rate of the young stars, which increased
  $\sim300$ Myr ago relative to the surrounding regions, is
  $(6.3^{+2.5}_{-1.5})\times10^{-5}$ \msol{} yr$^{-1}$
  kpc$^{-2}$. This implies a $\sim60\pm20$ Gyr depletion time of the
  \hi{} warp, similar to the timescales calculated for the outer \hi{}
  disks of nearby spiral galaxies.  While some stars associated with
  the warp fall into the asymptotic giant branch (AGB) region of the
  color magnitude diagram, where stars could be as old as 1 Gyr,
  further investigation suggests that they may be interlopers rather
  than real AGB stars.  We discuss the implications of these age
  constraints for the formation of \hi{} warps, and the gas fueling
  of disk galaxies.
\end{abstract}


\section{Introduction}
\setcounter{footnote}{6}

The neutral hydrogen (\hi) gaseous disks of many nearby disk galaxies
are warped. From a sample of 26 edge-on galaxies, \cite{gar02}
postulated that all \hi{} disks that extend beyond their optical
counterpart are warped. Unlike stellar warps, the onset of the \hi{}
warp is often abrupt, discontinuous, and in edge-on systems coincides
with the break in the exponential disk profile
\citep{bri90,van07}. However, the mechanisms forming these distortions
are still unclear \citep[for review, see][]{kru11}.

One possible explanation is that the warp is a manifestation of
ongoing gas accretion \citep[e.g.,][]{mac06,spa10}. Many galaxies show
possible signs of accretion in their \hi{} outskirts in the form of
bridges to companions, asymmetric density or velocity fields, and gas
clouds with anomalous velocities \citep[for review
see][]{san08}. However, the combined contribution of these phenomena
is unable to support the current star formation rates (SFRs) found in
the local universe\footnote{Although, as noted by \cite{han06}, such
  estimates of star-formation-rate densities carry large systematic
  uncertainties.} \citep{Tins80}.  Indeed, providing sufficient gas to
fuel star formation (SF) remains a key issue in $\Lambda$CDM
cosmology. Recent theoretical work suggests that for galaxies up to $L$*
in size, this gas is primarily accreted as cold, unshocked gas
\citep[e.g.,][]{dek06}. For larger systems, the quiescent cooling of
shock-heated gas after the last major merger plays the dominant role
\citep[e.g.,][]{bro04,rob06}. At early times ($z>2$), colder gas is
able to partially penetrate this hot halo through filaments. Although,
whether this gas remains cold or is shock heated before accretion is
still debated \citep[e.g.,][]{ker05,bro09,nel13}. If gas warps are a
sign of recent gas accretion, then they must play a significant role
in galaxy formation and evolution \citep[e.g.,][]{ost89,bin92,jia99}.

However, disk warps may be formed by mechanisms unrelated to accretion
\citep[e.g.,][]{bin78,hun69}. For instance, dynamical studies have
also reproduced warps purely through gravitational effects. These
torques may be due to a misalignment between the angular momentum of
the disk and dark-matter halo \citep{deb99} or by tidal interactions
with nearby galaxies \citep[e.g.,][]{hun69}, satellites
\citep[e.g.,][]{wei06}, or dark-matter substructure
\citep{kaz08}. Non-gravitational torques are also possible, e.g., from
passage through the intergalactic medium \citep{KahnWolt59} or
intergalactic magnetic fields \citep{bat90}. Although, such
magnetohydrodynamical processes may not be strong enough to cause the
observed warp angles \citep{bin92,bin00}.

Placing constraints on the age of \hi{} warps, and so determining how
long they remain coherent, may help determine their origin
\citep[e.g.,][]{ros11}. Hence, we need to identify the associated
stellar populations, which will encode a record of the formation of
the structure. SF beyond the optical disk was first inferred in the
outer disks of nearby galaxies via the detection of faint H{\sc ii}
regions \citep{fer98a,fer98b}. This extreme SF was subsequently
confirmed by observations of outer-disk UV emission
\citep{ThilBian07}. However, as discussed in \cite{dej08}, the
inherent degeneracies between stellar age, metallicity, and reddening
due to dust preclude a meaningful measurement of the stellar content
of these outer regions from integrated-light studies. Studies of
resolved stellar populations mitigate many of these problems and allow
us to probe the extremely low stellar densities found in the outer
envelopes of nearby galaxies \citep[e.g.,][]{rad12}.

In this paper, we present the first study of the resolved stellar
populations located within a distinct \hi{} warp. We use imaging from
the \textit{Hubble Space Telescope} ($HST$) Advanced Camera for
Surveys (ACS) of \n4565. This edge-on Sb galaxy resides at a distance
of 11.9$\pm$0.3 Mpc \citep{ghosts} and is believed to host a bar and
pseudobulge \citep{kor10}. In Section~\ref{sec:warp}, we discuss the
properties of the pronounced gas warp in this system. In
Section~\ref{sec:photometry}, we describe the reduction of these data,
which we then use in Section~\ref{sec:stars} to study the stellar
component of the warp. We further analyze the association of
asymptotic giant branch (AGB) stars with the warp in
Section~\ref{sec:agb}.  In Section~\ref{sec:mechanisms}, we discuss
the implications of this study for various formation mechanisms of
\hi{} warps before summarizing our findings in
Section~\ref{sec:summary}.

\section{The Warp of \n4565}
\label{sec:warp}

\begin{figure}
\begin{center}
\resizebox{85mm}{!}{\includegraphics{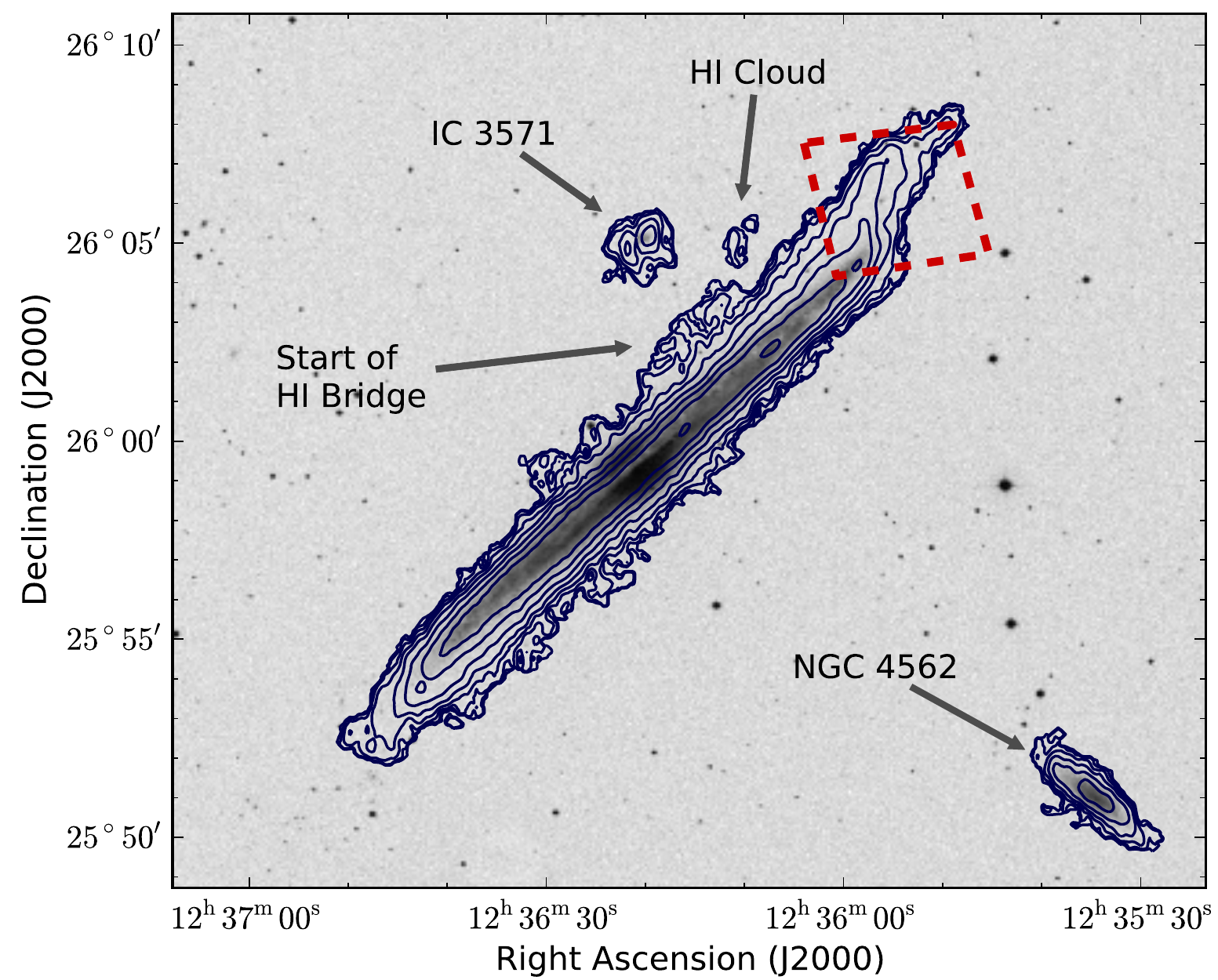}}
\caption{Distribution of neutral hydrogen \citep[\hi,][]{dah05}, shown
  as ten logarithmically spaced contours between column densities of
  $2\times10^{19}$ and $2\times10^{22}$ cm$^{-2}$, overlaid on a IIIaJ
  (blue) POSS-II image of \n4565. The location of the ACS field is
  shown as a dashed red box on top of the significant northwestern
  warp. The companion systems \n4562 and the smaller IC 3571 are seen
  at (R.A.,decl.) $\sim$
  ($12^{\mathrm{h}}35^{\mathrm{m}}34.8^{\mathrm{s}}$,
  $+25^{\mathrm{d}}51^{\mathrm{m}}00^{\mathrm{s}}$) and
  ($12^{\mathrm{h}}36^{\mathrm{m}}20.1^{\mathrm{s}}$,
  $+26^{\mathrm{d}}05^{\mathrm{m}}03^{\mathrm{s}}$),
  respectively. Part of a faint \hi{} bridge that connects \n4565 and
  IC 3571 is identified and extraplanar gas at $\sim$
  ($12^{\mathrm{h}}36^{\mathrm{m}}10^{\mathrm{s}}$, 26\degr05') is
  labeled as an \hi{} cloud.  }
\label{fig:ngc4565}
\end{center}
\end{figure}

As first noted by \cite{san76}, the \hi{} disk of \n4565 is strongly
warped, most notably on the northwestern (NW) side of the
galaxy. \cite{rup91} calculated the gas mass of the NW component of
the warp to be $\sim7\times10^8$\msol{}
\citep[c.f. $7.43\times10^9$\msol{} for the entire \n4565
system,][]{hea11}. However, with evidence for a mild warp along the
line of sight, \cite{rup91} suggested that the full warp is continuous
around the edge of the disk and encompasses a total mass of
$1.1\times10^9$\msol, or about $\sim15\%-20\%$ of the \hi{} mass of
the entire system. \cite{rup91} found this structure to be asymmetric,
in that the upward and downward bending modes of the warp are not
separated by 180\degr{} in azimuth but instead $\sim140$\degr. The
authors note, however, that this apparent lopsidedness may be due to a
lack of \hi{} in the southern component of the warp rather than an
inherent asymmetry. In Figure~\ref{fig:ngc4565}, the \hi{} data from
\cite{dah05} is overlaid as contours on a IIIaJ (blue) plate scan from
the Second Palomar Observatory Sky Survey \citep[POSS-II;][]{rei91}.

In order to present lower surface-brightness features, we have
reprocessed the \hi{} data, following a similar procedure to
\cite{walter08}. We first created a post-imaging mask by convolving
the data cube to a 30\arcsec{} $\times$ 30\arcsec{} resolution and
selecting all pixels with flux densities greater than twice the noise
in the convolved cube in three consecutive velocity channels. We then
edited this mask by hand to remove spurious signals such as sidelobes,
noise spikes, and other artifacts. The final mask was applied to the
original unconvolved data and a new zeroth moment map was generated by
summing the cleaned velocity channels.

Clearly evident in Figure~\ref{fig:ngc4565} is the abrupt upturn
($\sim45$\degr) of the NW warp. This structure begins at $\sim$7\farcm5
from the nucleus and coincides with a steep drop in both the density
and the rotational velocity of the \hi{} gas \citep{rup91}.  The
distinct morphologies of this warp gas and the in-plane gas led
\cite{van07} to conclude that the two components are distinct, with
the warp representing an accreting component. As evident in the
contours of Figure~\ref{fig:ngc4565}, the warp appears to turnover and
realign with the gas disk at greater distances, as first noted by
\cite{san83}.

In the optical, the upturn of the NW gaseous warp corresponds with a
truncation of the main stellar disk \citep{van79,wu02}. Internal to
this truncation, a mild warp is seen in the optical
\citep{van79,van81}. \cite{nae97} have also argued for a possible
extension of the optical disk beyond the truncation that coincides
with the NW \hi{} warp. In their $V$-band observations with the
2.56 m Nordic Optical Telescope, they measured a surface brightness
of $\mu_v\approx27$ mag arcsec$^{-2}$ for this optical
structure. However, they could not rule out this feature as a
background galaxy.

Also evident in Figure~\ref{fig:ngc4565} are the two companion
systems. To the southwest is \n4562, for which \cite{hea11} find a
regularly rotating \hi{} disk with a total \hi{} mass of
$1.82\times10^8$\msol{}. To the north is IC 3571
\citep[$M_{\mathrm{H}\,{\mbox{\tiny{I}}}}=4.18\times10^7$\msol;][]{hea11},
which shows evidence of interaction with \n4565. Specifically, a weak
\hi{} bridge is found between the disk of \n4565 and this companion
galaxy, which exhibits a smooth change in velocity between the two
systems \citep{van05,hea11}. The initial extension of this bridge from
\n4565 can be seen as an extended shelf in Figure~\ref{fig:ngc4565} at
(R.A., decl.) $\sim$ ($12^{\mathrm{h}}36^{\mathrm{m}}15^{\mathrm{s}}$,
26\degr03\arcmin). We also find evidence in Figure~\ref{fig:ngc4565}
for a smaller overdensity, or cloud, of \hi{} lying above the disk at
$\sim$ --($12^{\mathrm{h}}36^{\mathrm{m}}10^{\mathrm{s}}$,
26\degr05\arcmin) as also seen by \cite{zsc12}. This overdensity
appears connected to the bridge between \n4565 and IC 3571.

\section{$HST$ ACS Photometry}
\label{sec:photometry}

\begin{figure*}
\begin{center}
\resizebox{172mm}{!}{\includegraphics{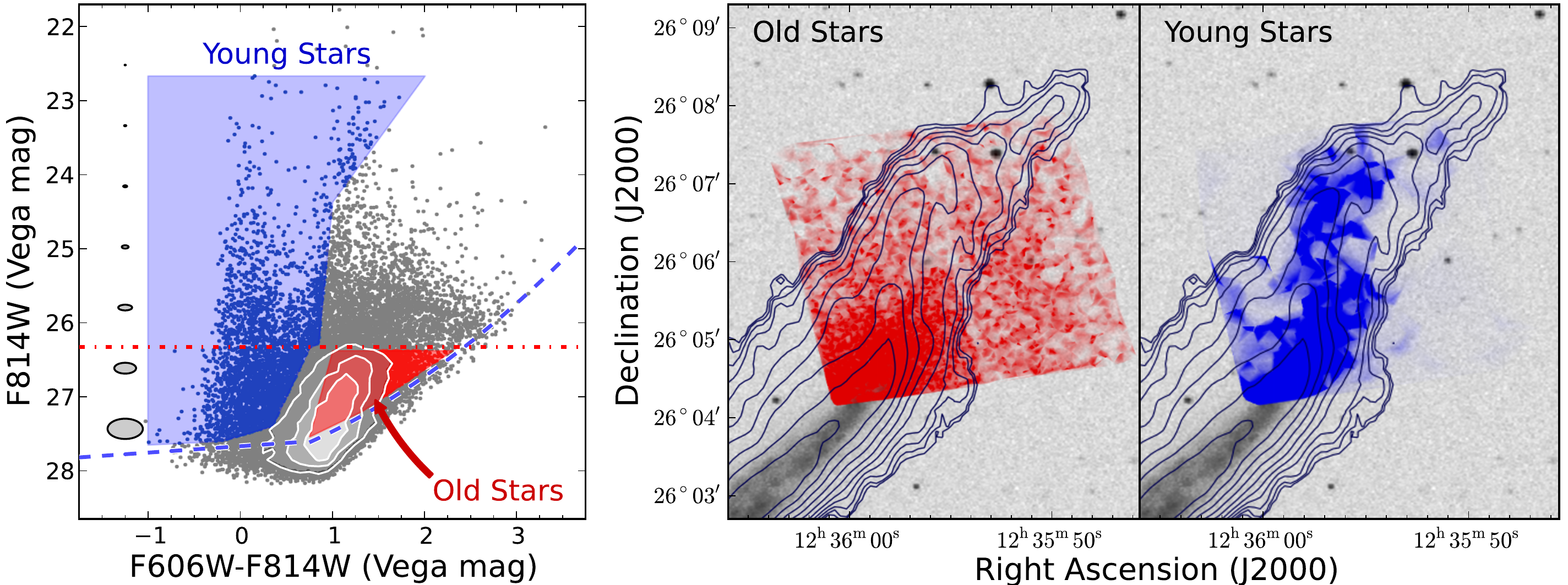}}
\caption{CMD of the stars in the ACS field is shown in the leftmost
  panel. Filled contours are used at densities greater than 50 stars
  in a 0.1 mag $\times$ 0.1 mag bin in F814W and color, with
  consecutive contours displayed at 100, 150, and 250 stars per 0.01
  mag$^2$. The TRGB is indicated by a red dot-dashed line, while the
  50\% completeness limit measured from the artificial star tests is
  marked by a blue dashed line. Old RGB stars are indicated in red,
  while young MS and HeB branches are colored blue. Ellipses indicate
  the photometric uncertainties reported by \dol, while the arrow
  indicates the direction of foreground reddening, which has been
  corrected for in these data. The spatial densities of these old and
  young stars are shown in the middle and right panel,
  respectively. The densities are plotted as Voronoi tessellations on
  top of an $Rc$-band POSS image with \hi{} gas column densities
  spaced logarithmically by 0.2 dex between $5\times10^{19}$ and
  $5\times10^{21}$ cm$^{-2}$. A clear correlation between the gas and
  the young stars is seen, but no such coincidence is seen with the
  older stars.}
\label{fig:cmd}
\end{center}
\end{figure*}

The ACS field covers the extent of the \hi{} NW warp, as indicated in
Figure~\ref{fig:ngc4565} by a dashed box. The field was observed with
exposures of 8266 s and 7340 s using the F606W and F814W filters,
respectively ($HST$ GO program 12196). Each exposure was dithered to
aid cosmic-ray rejection and to cover the chip gap. However, in the
subsequent analysis, we removed detections from the chip gap region
where the detection efficiency is less due to the shorter mean
exposure time. We followed the same procedures for the data reduction
as detailed by \cite{ghosts}. This involved using \textsc{SExtractor}
\citep{ber96} to mask out resolved sources and the ACS module of
\dol{} \citep{dol} to identify stars and measure their photometry. We
also used the same crowded-field selection parameters selected by
\cite{ghosts} to remove the bulk of unresolved background
galaxies. Finally, we generated approximately 250,000 artificial stars
that mimicked the color and magnitude distribution of the observed
photometry. These artificial stars were added to the ACS image and
subsequently passed individually through the same photometry pipeline
and selection criteria. By measuring the recovery rate of these stars,
we can assess incompleteness in the real data at fainter magnitudes
and measure the effects of stellar crowding in the disk. We found that
at magnitudes of F814W$<27.6$ mag, more than 50\% of these stars are
recovered. At this magnitude \dol{}, reports the typical uncertainty
as 0.07 mag. The resulting color-magnitude diagram (CMD) of the
stellar detections is shown in the leftmost panel of
Figure~\ref{fig:cmd}.

By analyzing synthetic CMDs similar to Figure~\ref{fig:cmd}, we can
identify and date several distinct regions corresponding to different
stages of stellar evolution. Such models depend critically on
metallicity, which in outer disks is found to range from
$\textrm{[M/H]}=-1$ to $-0.5$ \citep{gil07,bre09}. If the warp is
formed from infalling material, we may expect a particularly
metal-poor composition, hence we generated these CMDs with [M/H]
$=-1$.

Using our model CMDs, we identified the region labeled `Old Stars' in
Figure~\ref{fig:cmd} to be principally composed of red-giant-branch
(RGB) stars, which at 1--10 Gyr are the oldest stellar population
present as well as the most abundant. These stars have a well-defined
maximum luminosity, which corresponds with the sharp cutoff known as
the tip of the red giant branch (TRGB). Immediately above these TRGB
stars is an overdensity of the intermediate aged (1--5 Gyr) AGB stars.

The plume at F606W--F814W$=$1 corresponds to the red helium-burning
(HeB) branch. These stars, which are undergoing core helium burning,
are around 30--600 Myr old. The leftmost plume at F606W--F814W$=$0
traces the main-sequence stars (MS), which are typically $<$100
Myr. As shown by the shaded region of the CMD, we have binned these
two features together to represent young stars. In the middle and
rightmost panel of Figure~\ref{fig:cmd}, the spatial density of these
old and young stars, shown as Voronoi tessellations, are compared with
the reprocessed \hi{} gas map as indicated by the contours. A clear
correlation with the gas warp is seen with the young stars but not
with the old stars. After correcting for completeness, the mean
density of these young stars ($22.7<$F814W$<27.6$) in the warp is
$(2.8\pm0.1)\times10^{-2}$ stars arcsec$^{-2}$. This is approximately
four times less than the mean density of the old stars
($26.4<$F814W$<27.5$) at $(1.25\pm0.03)\times10^{-1}$ stars
arcsec$^{-2}$.

\section{The stellar component of the warp}
\label{sec:stars}

\begin{figure*}
\begin{center}
\resizebox{175mm}{!}{\includegraphics{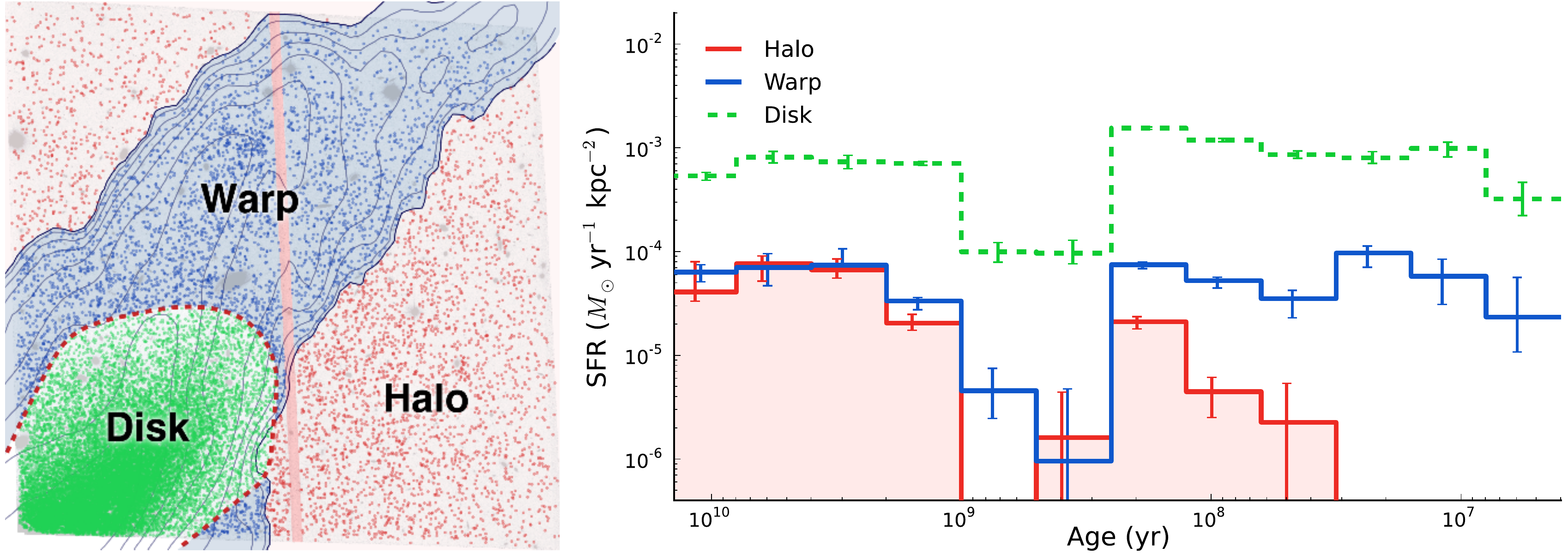}}
\caption{Left panel shows all the stellar detections superimposed on
  the F814W image. The \hi{} column density is plotted as contours
  spaced logarithmically by 0.2 dex between $5\times10^{19}$ and
  $5\times10^{21}$ cm$^{-2}$. The smoothed dashed line bounds the
  region containing stellar detections greater than 0.5 stars
  arcsec$^{-2}$, which we define as the disk population. The warp
  population is composed of stars residing in regions with coincident
  \hi{} flux $>5\times10^{19}$ cm$^{-2}$ and that are not part of the
  disk population. Stars in the remaining area are identified as the
  halo population. We mask out all stellar detections in the chip gap
  region, which is marked by a vertical strip. The right panel plots
  the SFHs of these separate populations in logarithmic age bins of
  width 0.3 dex. Error bars indicate uncertainties based on a Markov
  Chain Monte Carlo exploration of the probability space. Note that
  the uncertainties in the underlying stellar models are not included
  as we are only interested here in the relative difference of the
  SFHs rather than the absolute calibration. A pronounced difference
  is seen between the star formation rates in the warp and halo at
  ages $<300$ Myr.}
\label{fig:sfh}
\end{center}
\end{figure*}

To study the star formation history (SFH) of the warp, we first
defined the warp as the region of the ACS image with coincident \hi{}
detection ($>5\times10^{19}$ cm$^{-2}$). We excluded any detections
from the main disk, which we defined as the region in the southeast
corner of the ACS field bound by a stellar density of 0.5 stars
arcsec$^{-2}$. The remaining area is attributed to the stellar
halo. To measure the SFH of these regions, we used the software
package {\textsc{match}} \citep{dol02}, which fits the observed CMD
with synthetic stellar populations. These populations were generated
using stellar evolution models and a model of completeness and
photometric errors, which is computed from the artificial star tests
described in Section~\ref{sec:photometry}. For this analysis, we used
the stellar evolution models of \cite{mar08} with the updated AGB
tracks of \cite{gir10} and adopted both a \cite{kro01} initial mass
function (IMF) and a binary fraction of 40\%. These synthetic
populations were distributed across a 36$\times$25 grid in
$\log(\rm{age})$-[$Z$] space with bin widths of 0.1 dex. The distance
modulus was set to $m-M=30.38$ \citep[as measured from the
TRGB,][]{ghosts}, the extinction was set to $A_V=0.05$ \citep{sch98},
and the chemical enrichment history was required to increase in
metallicity with time. {\textsc{match}} also allows for up to 0.5 mag
of differential reddening for stars younger than 40 Myr, decreasing to
0 mag for stars older than 100 Myr. With this setup, we measured the
SFR in fixed logarithmic age bins of width 0.3 dex. To determine the
relative uncertainties in SFR between different regions
{\textsc{match}} employs a Markov Chain Monte Carlo approach to sample
the probability space around the original data.

The regions, and their respective SFHs, are shown in
Figure~\ref{fig:sfh}. For ages older than $\sim300$ Myr, the SFRs of
the halo and warp are remarkably similar. This suggests that the old
stars found in the warp are likely contamination from the halo and are
not directly associated with the warp. However, a clear difference
between the halo and warp populations is seen at younger
ages. Specifically, the SFR of the halo rapidly drops off and is
undefined at ages $<20$ Myr while stars in the warp sustain a steady
SFR of $(6.3^{+2.5}_{-1.5})\times10^{-5}$ \msol{} yr$^{-1}$ kpc$^{-2}$
over the last 250 Myr. Following \cite{dol12}, this measurement
includes uncertainties in the stellar evolutionary models by
incorporating shifts in both luminosity and temperature when computing
the Monte Carlo realizations.

Figure~\ref{fig:sfh} also shows a dearth of SF between 1 Gyr and 300
Myr ago. The subsequent rapid increase in SF across all components
after this period may indicate a significant event that triggered both
the SF and the onset of the warp. However, this feature may also be
attributed to a lack of sensitivity at these intermediate
timescales. Such stellar ages encompass AGB stars (discussed in
Section~\ref{sec:agb}) and faint HeB stars that are at the detection
limit of the ACS exposure.

{\textsc{match}} accounts for interstellar extinction in the fits by
allowing for up to 0.5 mag of differential reddening for the very
young stars found in dusty stellar nurseries. However, as an extreme
case, we may ask how the SFH results would change if interstellar dust
affected all populations. To estimate this level of extinction, we
used the relation between \hi{} column density and dust extinction as
measured locally by \cite{boh78}. Given that the average projected
surface density of \hi{} in the warp is $4\times10^{6}$
\msol~kpc$^{-2}$, this relation yields a total extinction of
$A_V\sim0.26$ mag. If the stars are embedded in the gas, then half
this value may be assumed as the typical stellar extinction. However,
we note that such a calculation is likely to overestimate dust
extinction in low-metallicity environments such as those expected to
be found in the warp \citep{ler11}. To test the impact on the SFH from
a reddening of this order of magnitude, we allowed {\textsc{match}} to
fit up to 0.5 mag of differential extinction for all stellar
populations. The resulting fits were consistent with the best-fit SFHs
presented in Figure~\ref{fig:sfh} given the {\textsc{match}} reported
total uncertainties. We thus conclude that interstellar reddening has
little effect on the SFHs.

The projected surface density of \hi{} in the warp is comparable to
the typical density of \hi{} outer disks in nearby spiral galaxies
\citep{bigiel10b}.  Using the warp SFR as calculated by
{\textsc{match}}, we infer a depletion time of $\sim60\pm20$ Gyr. This
is also similar to the $\sim$100 Gyr depletion times of quiescent
outer disks as found by \cite{bigiel10b} using far-UV (FUV) emission
as a proxy for SFR. This suggests that the same factors setting SF
efficiency (defined as the inverse of the depletion time) in galaxy
outer disks are also operating in the \n4565 \hi{} warp.

\subsection{Stellar Populations}

\begin{figure*}
\begin{center}
\resizebox{175mm}{!}{\includegraphics{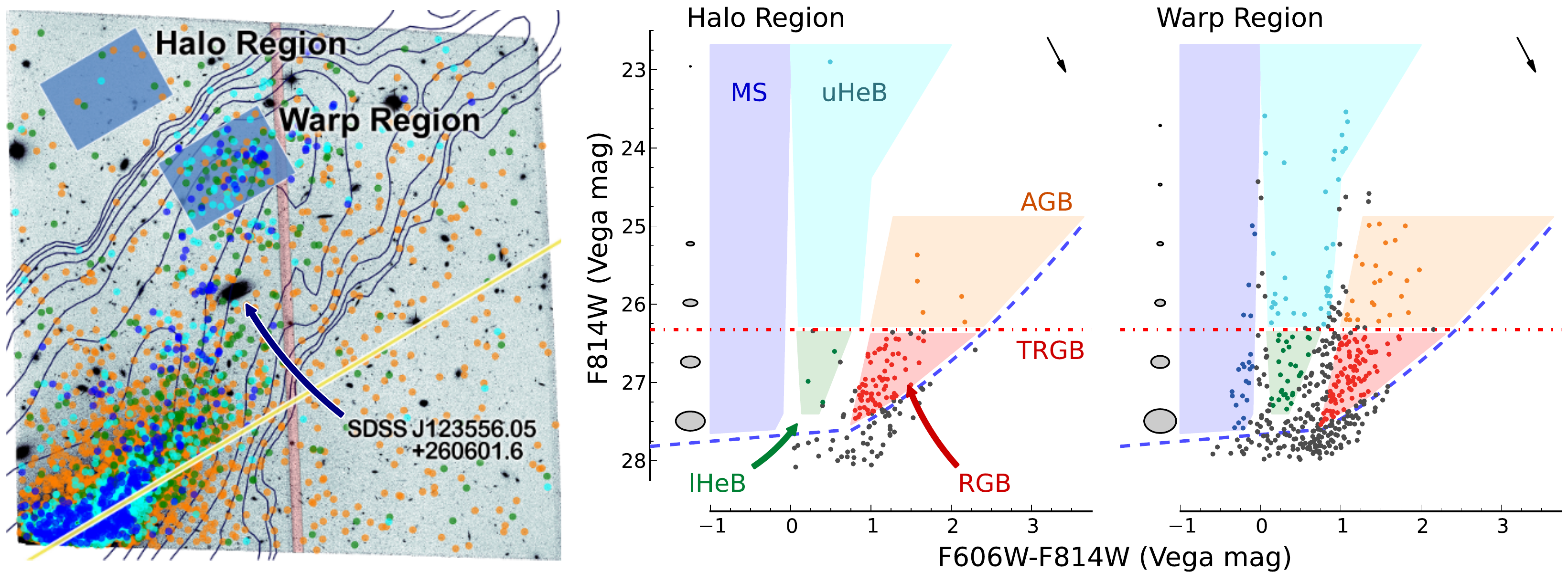}}
\caption{Stellar content of the warp. The left panel shows stellar
  detections superimposed on the $HST$ ACS F814W image, with \hi{}
  column densities plotted as contours spaced by 0.2 dex between
  $5\times10^{19}$ and $5\times10^{21}$ cm$^{-2}$. The chip gap, which
  we remove from our analysis, is indicated by a nearly vertical red
  strip. The sloping yellow line indicates the plane of the disk
  inferred from the stellar density in the lower left of the
  image. The colors of the points represent the distinct stellar
  populations: MS, lHeB, uHeB and AGB stars as labeled in the CMD in
  the middle panel. RGB stars and detections that do not lie in one of
  these age bins make up the majority of detections, but are not shown
  in the left panel for clarity. The background galaxy SDSS
  J123556.05+260601.6, which will contaminate UV studies of the warp,
  is labeled. Two rectangular regions inside and outside of the warp
  are indicated by shaded boxes. Detections within these regions are
  displayed as CMDs in the middle and rightmost panel. Regions
  denoting different stellar populations, as described in the text,
  are shaded. The 50\% recovery rate as measured from the artificial
  star tests is indicated by a dashed blue line, and arrows indicate
  the direction of the corrected foreground reddening. The CMD in the
  warp region shows an abundance of young stars not seen in the halo
  region.}
\label{fig:stellarwarp}
\end{center}
\end{figure*}

\begin{figure*}
\begin{center}
\resizebox{154mm}{!}{\includegraphics{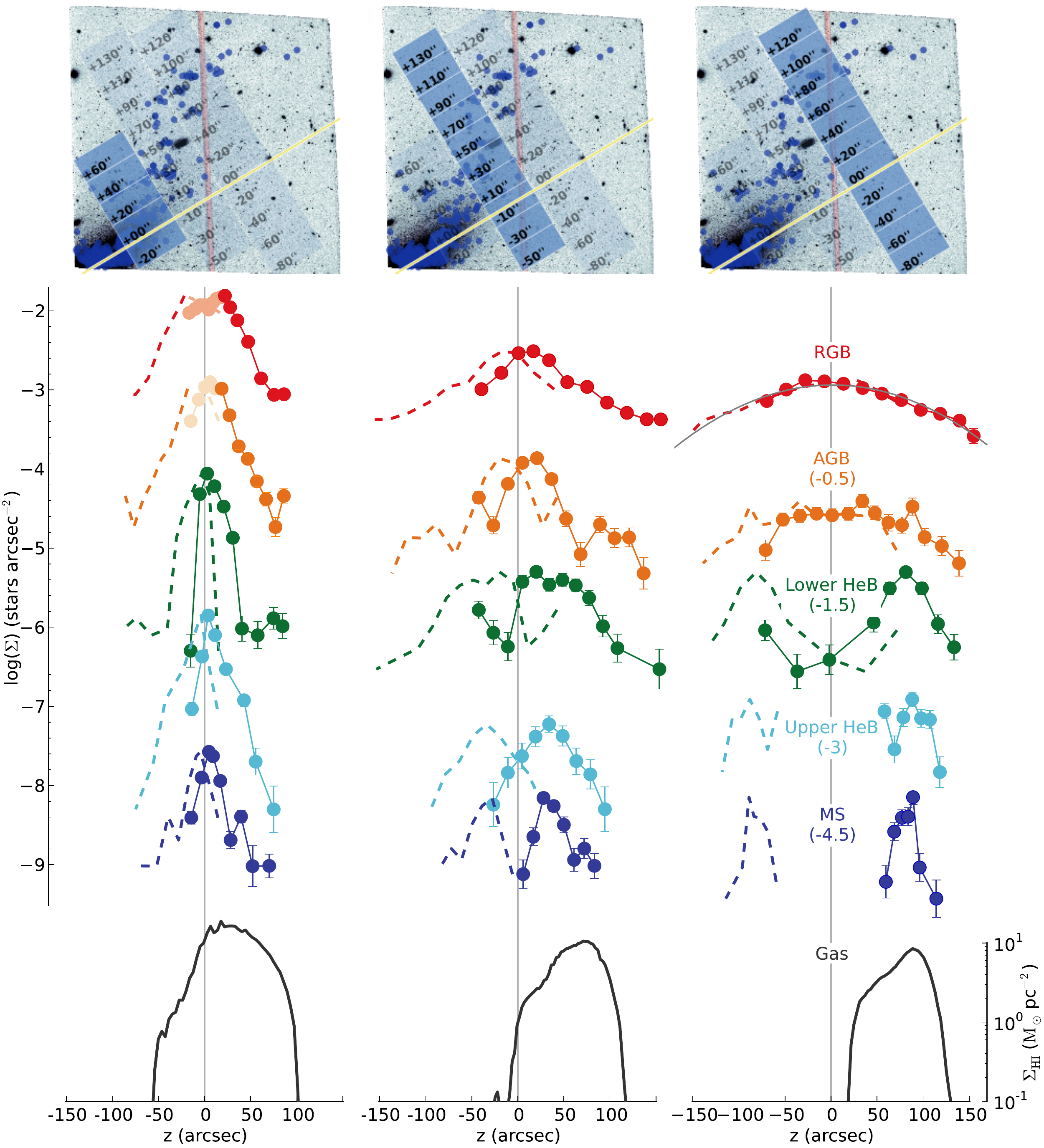}}
\caption{Correlation of the stars and gas across the plane of the disk
  in the \n4565 observations. The shaded regions in the top panels
  indicate the cuts perpendicular to the plane of the disk, which is
  marked by a solid yellow line. Blue dots indicate the distribution
  of MS stars in the ACS region. The chip gap is shown as a vertical
  red stripe and any detections falling in this region are
  removed. The first cut (left column) covers the start of the warp
  from the main disk, while the third (right panel) cuts across the
  full warp. The second cut (middle panel) is intermediate. Below
  these plots are the stellar densities along the cuts of the old RGB
  stars (top red points), AGB, lHeB, uHeB, and MS stars. The AGB,
  lHeB, uHeB, and MS profiles have been offset by $-0.5$, $-1.5$,
  $-3$, and $-4.5$ dex for clarity. The \hi{} gas density along the
  strips is shown as a gray solid line at the bottom of each
  column. The stellar profiles are reflected around the disk plane and
  plotted as dashed lines. A clear correlation between the young stars
  and gas is seen in the right panel. The corresponding distribution
  of old stars instead shows a symmetric distribution around the disk
  plane, as indicated by a second order polynomial fit to the RGB
  profile, which is plotted as a thin gray line.}
\label{fig:cross}
\end{center}
\end{figure*}

Due to the shallow depth of the observations, the SFHs derived in
Figure~\ref{fig:sfh} may lack sensitivity at intermediate ages. Hence,
to further examine the stellar content of the \hi{} warp, we
investigated the CMD of individual detections. Specifically, we
examined detections in a region of the warp with a pronounced
overdensity of stars. This includes a massive star cluster lying at
(R.A., decl.) $\sim$ ($12^{\mathrm{h}}35^{\mathrm{m}}56^{\mathrm{s}}$,
$+26^{\circ}06$\arcmin54\farcs5), which is misidentified in the Sloan
Digital Sky Survey \citep[SDSS;][]{yor00} as a galaxy. By focusing on
this overdensity, we minimize contamination from projected halo stars
and so maximize the difference between warp and halo populations. This
region also avoids the large background galaxy SDSS
J123556.05+260601.6, which is resolved in the ACS image. This galaxy
can be seen in FUV and near-UV (NUV) imaging from the \textit{Galaxy
  Evolution Explorer} \citep{mar05} archive. Although there is some
evidence of UV emission from the surrounding warp, the flux in the
region is mostly attributed to this galaxy, which thus hampers studies
of the UV warp.

We defined a new halo region, which is equal in size ($\sim$1300
arcsec$^2$) after adjusting for the area removed from the warp region
due to the chip gap. For greater age fidelity, we split the CMDs into
the distinct stellar populations discussed in
Section~\ref{sec:photometry}. Based on the stellar isochrones used to
construct the SFHs derived earlier, we further split the
helium-burning branches into an upper helium-burning branch (uHeB),
which corresponds to an age range of 30--200 Myr, and a lower
helium-burning branch (lHeB) of 100--600
Myr. Figure~\ref{fig:stellarwarp} plots the new warp and halo regions,
and their resulting CMDs.

Clearly evident in the CMD of the overdensity in the warp region is an
abundance of younger stars (MS, uHeB and lHeB) that are not present in
the halo region. This suggests, as also determined by the SFHs, that
SF is ongoing in the warp. However, the density of AGB and
RGB stars also appears higher in the warp region. This is likely due
to an older, flattened envelope surrounding the main disk, such as a
thick disk or inner stellar halo. Such a structure would yield higher
densities of old stars closer to the disk axis. To place age
constraints on the \hi{} warp, we thus need to assess the correlation
of these older stars with the warp.

In Figure~\ref{fig:cross}, we define three strips running
perpendicular to the disk plane. Along each of these strips, we plot
separate stellar density profiles for the RGB, AGB, lHeB, uHeB, and MS
detections, as well as the surface density of \hi{} gas. As shown by
\cite{ghosts}, these stellar densities are equivalent to surface
brightnesses. In the crowded region of the disk, detections are lost
due to the overlapping point-spread functions of the stars. Although
this effect can be partially corrected by artificial star tests, at
extreme densities the correction factor can become too large and
therefore unreliable. We thus excluded the high density regions within
20\arcsec of the disk plane as indicated in the leftmost panels by the
lightly shaded points. We also excluded the chip gap from the analysis
as the total exposure time in this region is shorter than the rest of
the field. This results in a brighter completeness limit, which would
affect detections in the fainter CMD selection bins. To illustrate the
symmetry of the profiles, dashed lines in the figure correspond to
reflections of the profiles around the disk axis.

The leftmost panel of Figure~\ref{fig:cross} lies largely within the
dominant main disk, and so all populations are found to approximately
correspond with the gas distribution. We note that this distribution
is slightly asymmetrical as the main stellar disk itself gradually
warps upward. The middle plot shows the transition from the disk to
warp populations, which is described by a gradual increase in the
asymmetry of the populations with decreasing age. However, the
rightmost panel, which covers the strip in the warp, shows a clear
discrepancy between both the RGB and AGB profiles with the younger
populations. The RGB stars are symmetrically distributed around the
disk axis, as would be expected from a very extended thick-disk-like
component or an oblate stellar halo. The AGB population appears to be
similarly distributed except for a small overdensity coincident with
the peak of the gas distribution. However, the peaks of the asymmetric
lHeB, uHeB, and MS populations all strongly coincide with the peak of
the \hi profile. This again suggests that the majority of the RGB and
AGB stars are not associated directly with the \hi{} structure but
rather with an extended outer envelope that is coincident with the
warp when seen in projection. Future observations to the south of the
field, i.e., $<-50\arcsec$ from the disk plane, would help measure the
symmetry of this older component and so rule out coincidence with the
\hi{} warp.

Given the stochastic uncertainties of the profiles in
Figure~\ref{fig:cross}, a sufficiently small fraction of the old RGB
stars may be associated with the warp without producing any detectable
peak. However, as the bulk of detections in the warp are RGB stars
(see Figure~\ref{fig:stellarwarp}), this small fraction may still be
significant. In order to date the warp using stellar ages, we thus
need to assess our detection sensitivity to these old stars.

Using the SFH analysis presented in Section~\ref{sec:stars}, we
constructed a series of artificial CMDs with increasingly inflated
SFRs at ages older than 1 Gyr. The excess RGB stars generated from
these higher SFRs were then spatially distributed in the warp
following the density of young stars. Finally, the RGB profile from
Figure~\ref{fig:cross} was recomputed and the signal-to-noise of any
peak near the warp was measured. Using this assessment, we found that
in order to produce a statistically significant peak, the old SFR in
the warp needed to be increased by an additional $\sim4\times10^{-5}$
\msol{}~kpc$^{-2}$~yr$^{-1}$ relative to the SFR in the halo. This is
comparable to the SFR seen in the warp today. Hence, the lack of a
peak in the RGB distribution of Figure~\ref{fig:cross} does not
preclude the existence of older SF associated with the warp. However,
we note that such SF would likely be seen in Figure~\ref{fig:sfh}, as
the SFH analysis is sensitive to much lower SFRs in the warp relative
to the halo at these stellar ages ($<1\times10^{-5}$
\msol{}~kpc$^{-2}$~yr$^{-1}$).

\section{AGB Stars}
\label{sec:agb}

\begin{figure*}
\begin{center}
  \resizebox{175mm}{!}{\includegraphics{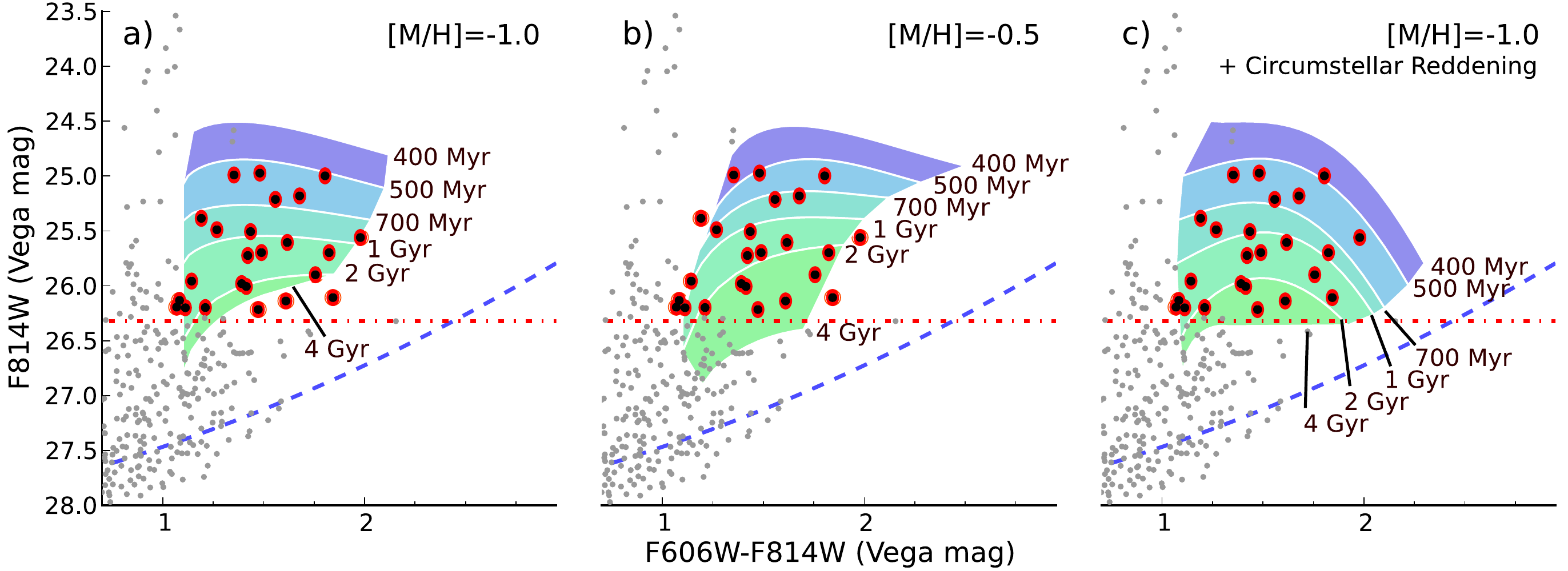}}
  \caption{CMDs of the AGB stars in the stellar-overdensity region
    found in the warp of \n4565. Approximate upper-age limits to these
    stars are indicated by colored strips for different theoretical
    models. Stars identified in Figure~\ref{fig:stellarwarp} as AGBs
    by their color and magnitude are indicated by heavy black dots
    with a red border. The age bins are generated from the AGB
    evolutionary models of \cite{mar08} and \cite{gir10} and include
    the thermally pulsating stage of the AGB sequence. Panel (a) uses
    the default parameters with no circumstellar dust, an interstellar
    extinction of $A_V=0.042$ mag, and a metallicity of $Z=0.0019$
    ([M/H] = $-1$). Panel (b) increases the metallicity to $Z=0.006$
    ([M/H] = $-0.5$), while panel (c) uses the original metallicity
    but includes the effects of circumstellar reddening. Panels (a)
    and (b) show an increase in age with decreasing F814W magnitude,
    with metallicity having little effect on ages less than $\sim1$
    Gyr. Including circumstellar reddening increases the dependence of
    stellar age on color.}
\label{fig:agb}
\end{center}
\end{figure*}

\begin{figure*}
\begin{center}
  \resizebox{150mm}{!}{\includegraphics{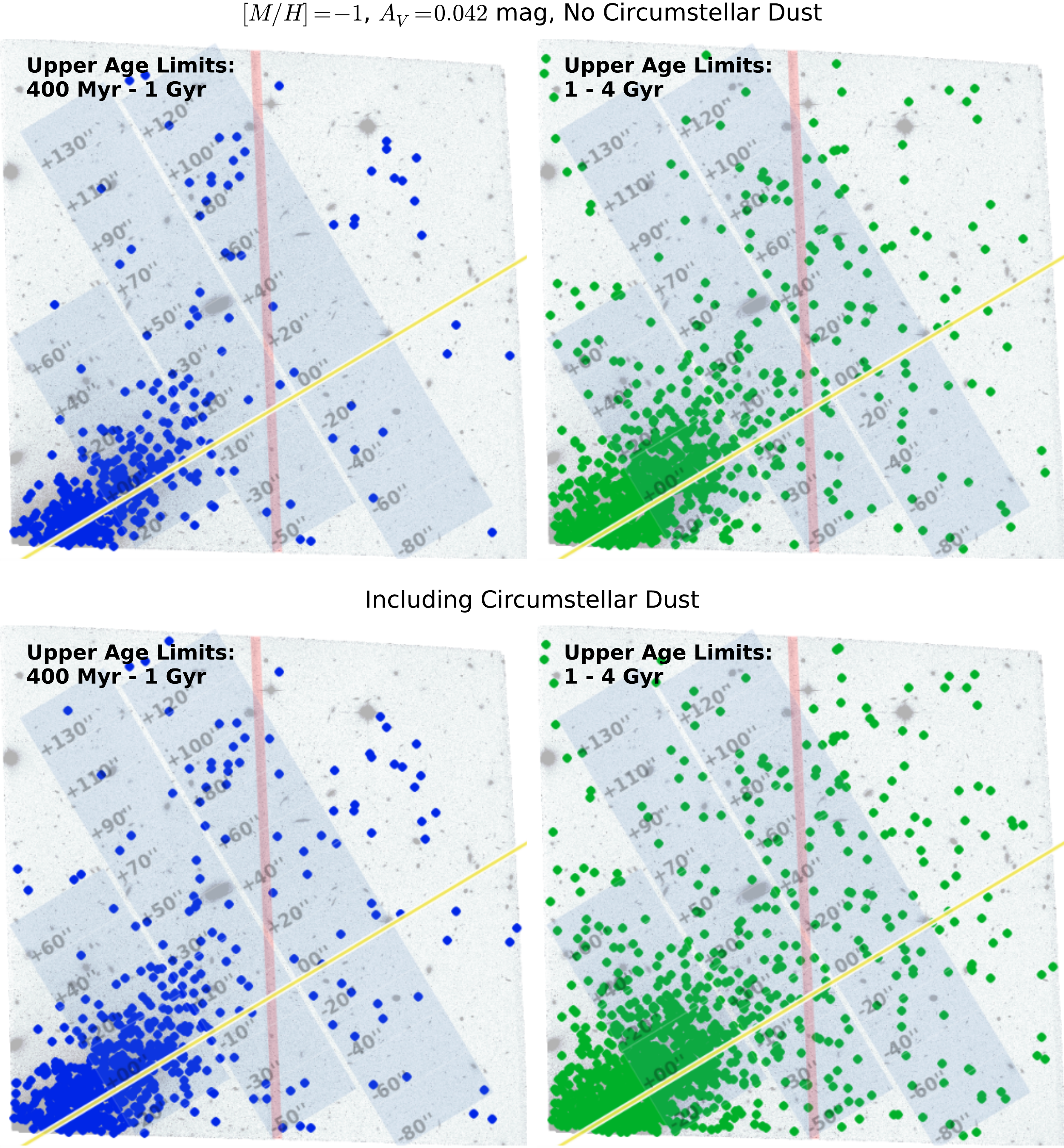}}
  \caption{Spatial distribution of stars in the ACS field that are
    brighter than the TRGB and fall in the upper-limit age bands
    defined in Figure~\ref{fig:agb}. The leftmost panels plot only the
    stars that fall in the 400 Myr to 1 Gyr bands, while the rightmost
    panels use the 1-4 Gyr bands. The upper two plots use our baseline
    stellar evolutionary models from \cite{mar08} and \cite{gir10} (no
    circumstellar dust, an interstellar extinction of $A_V=0.042$ mag,
    and a metallicity of [M/H] = $-1$). In these panels, the younger
    ($<1$ Gyr) stars show a correlation with the warp that is less
    apparent in the older stars. The bottom two panels repeat the
    process but use the evolutionary models with circumstellar
    dust. The apparent correlation of the young stars with the warp is
    diminished, suggesting that the circumstellar dust models are not
    as good a match to these data.}
\label{fig:agbspatial}
\end{center}
\end{figure*}

In the outermost profile of Figure~\ref{fig:cross}, a small excess of
AGB stars is seen in the region of the warp ($\sim+90$\arcsec). As
previously noted, AGB stars are typically a few Gyr old and so may
represent the oldest component of the warp. In Figure~\ref{fig:agb},
we replot the CMD of the warp stars in the stellar overdensity defined
in Section~\ref{sec:stars}, which intersects with the third strip of
Figure~\ref{fig:cross}. Overplotted in each panel are binned regions
of constant age using different assumptions. The underlying isochrones
are extracted from the models of \cite{mar08} using mass-loss
corrections to the AGB tracks from \cite{gir10}. However, we note that
as stars evolve upward through the AGB region these bands should be
treated as upper-age limits. The bands also average over the effects
of thermal pulsations. Typically the effects of these pulsations are
smaller than the vertical (F814W) extent of each age bin. For our
baseline instance, we use a \cite{cha01} log-normal IMF, a total
extinction of $A_V=0.042$ mag with no contribution from circumstellar
dust, and a metallicity of $Z=0.0019$ ([M/H] = $-1$). As previously
noted, this represents the lowest metallicities found in outer disks
\citep{gil07}. With this configuration (leftmost panel of
Figure~\ref{fig:agb}), the youngest AGB stars correspond to $\sim500$
Myr. The ages subsequently increase approximately linearly with F814W
magnitude until about 2 Gyr, when the effects of color become more
pronounced. AGB stars at the TRGB are typically older than 4 Gyr.

In the middle panel of Figure~\ref{fig:agb}, we dramatically increase
the metallicity of the isochrones to $Z=0.006$ ([M/H] = $-0.5$), as
found in the outer disks studied by \citep{bre09}. This has only a
minor effect on the evolutionary tracks for stars younger than 1 Gyr,
generally pushing stars toward slightly older ages. However, a much
larger effect is seen in the rightmost panel of Figure~\ref{fig:agb},
where we revert to our original metallicity ([M/H] = $-1$) but use
isochrones that include circumstellar reddening. To achieve this,
\cite{mar08} tracked the C/O ratio in their model stellar atmospheres
for stars that experienced significant mass loss. Depending on this
ratio, each star was then reddened using the AGB dust absorption
models of \cite{gro06}, either with a mix of 60\% aluminum oxide
(AlOx) and 40\% silicates for oxygen-rich stars, or 85\% amorphous
carbon (AMC) and 15\% silicon carbide (SiC) for carbon-rich
stars. With this addition, a strong dependence on color is now seen,
with large extinction values in F814W for redder colors.

We explored this age dependence on the spatial distribution of AGB
stars across the entire ACS field, as shown in
Figure~\ref{fig:agbspatial}. In this figure, we only plot stars
brighter than the TRGB that fall in the upper-age bands of
Figure~\ref{fig:agb}, specifically the 400 Myr to 1 Gyr bands (shown
in the left panels) and 1 Gyr to 4 Gyr bands (right panels). Thus not
all stars that fall in the classical AGB region are
shown. Furthermore, as the bands in each panel of Figure~\ref{fig:agb}
cover different regions of the CMD, selections based on these various
models will sample different subsets of the entire AGB
population. Using the age definitions from the evolutionary models
without circumstellar reddening (upper panels), we found a high
spatial correlation between the \hi{} warp and stars $<1$ Gyr. This
correlation is weaker for AGB stars $<4$ Gyr. Again, we note that as
these age definitions are upper limits, young stars ($<1$ Gyr) may
still contribute to the remaining correlation seen in the vicinity of
the gas warp in the upper-right panel. If we repeat this analysis with
the models including circumstellar extinction (bottom panels), the
correlation is less apparent for all age bins. Given that AGB stars
$<1$ Gyr coincide so well with the \hi{} warp only for the cases
without circumstellar reddening, we can reason a priori that these
evolutionary models are a better match to these data.

In Section~\ref{sec:stars}, we estimated based on the \hi{} gas
density that interstellar reddening from dust may account for up to
$A_V\sim0.13$ mag of extinction. Correcting for this would shift the
stars by $-0.04$ mag in color and $-0.08$ mag in F814W magnitude,
slightly decreasing the measured ages of the stars. However, we
disregarded such shifts in this analysis as they are smaller than the
typical systematic uncertainties in the models (e.g., the differences
between the left and middle panels of Figure~\ref{fig:agb}).

If these stars do indeed place an upper limit of $\sim1$ Gyr on the
age of the \hi{} warp structure, we need to assess the validity of
using AGB evolutionary models to tag their age. That is, we should
assess the likelihood that these stars are AGB stars based solely on
their position in the CMD.

\subsection{Significance}

\begin{figure*}
\begin{center}
\resizebox{165mm}{!}{\includegraphics{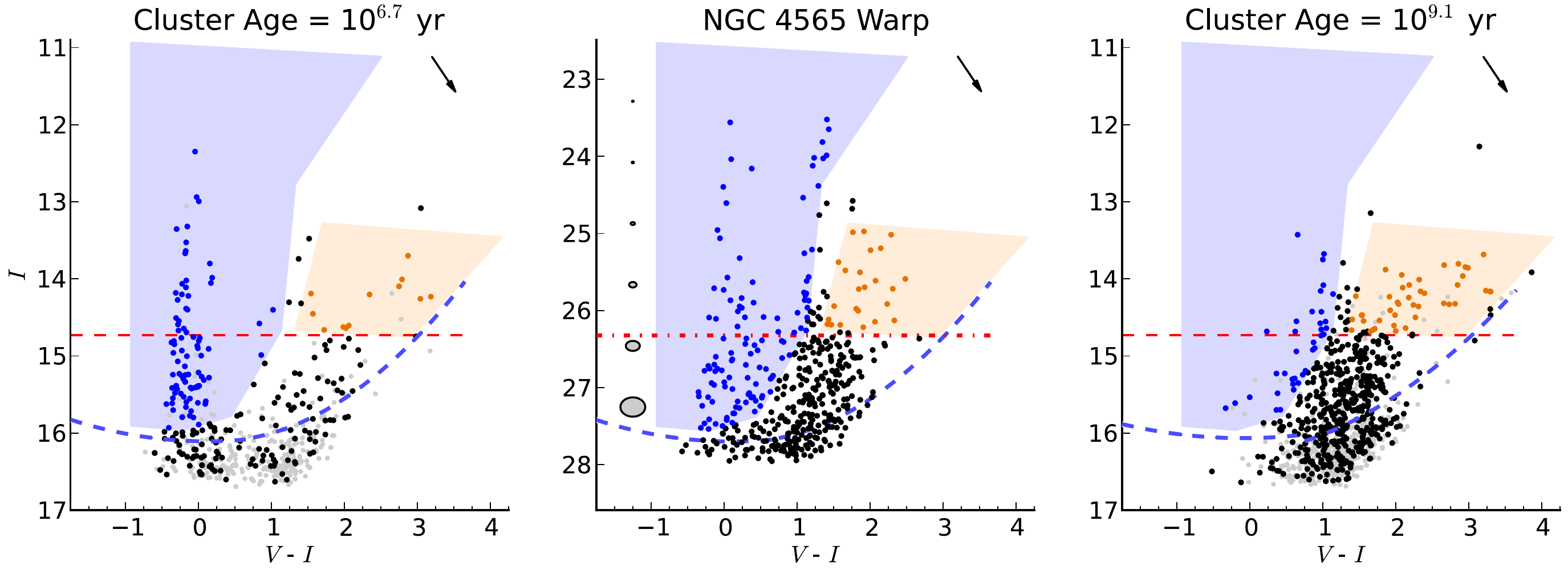}}
\caption{CMDs of the degraded LMC clusters and the stellar-overdensity
  region in the \n4565 warp plotted in the Johnson-Cousins system used
  by OGLE. Gray points indicate degraded stars that were discarded by
  the selection criteria used to cull the \n4565 warp photometry. The
  left panel plots the degraded data from the combined 34 LMC clusters
  in the youngest age bin. A clear overdensity of stars falling in the
  AGB region is seen even though the cluster is too young to contain
  AGB stars. The middle panel plots these data from the overdensity of
  the warp in \n4565, which shows that same overdensity. The right
  panel plots the degraded data from the 64 LMC clusters in the oldest
  age bin. The main sequence is no longer visible but a clear
  abundance of old AGB and RGB stars is seen.}
\label{fig:lmc}
\end{center}
\end{figure*}

We studied the dependency of stellar age on position in a CMD by using
data from the Optical Gravitational Lensing Experiment
\citep[OGLE;][]{uda97}. This survey provides photometry and ages of
approximately 600 star clusters in the Large Magellanic Cloud (LMC),
spanning ages from $\sim3$ Myr up to $\sim$1.2 Gyr \citep{pie00}. To
assess contamination from field stars in these systems, we also used
photometry from the complimentary OGLE maps of dense stellar regions
\citep{uda00}. To compare these data with \n4565, we transformed the
$HST$/ACS F606W and F814W photometry to $V$- and $I$-band photometry
using the conversions listed in \cite{sir05} and dereddened all
photometry using the extinction values published by \cite{sch98}.

We began by arranging the LMC clusters into nine age bins, spaced
logarithmically by 0.3 dex. We then binned the photometry of each
cluster into a Hess diagram at a resolution of 0.1 mag in color and
0.2 mag in the $I$-band equivalent. For each of these clusters,
\cite{pie00} defined a core radius, marked by a major drop in radial
stellar density, as well as a total cluster radius, calculated by
fitting profiles to the radial stellar density. We subsequently
defined a sky region for each system as the annulus that extended
outward from the total cluster radius and encompassed an area equal to
that of the cluster. Stars from the OGLE stellar maps were then
selected in this region by their R.A. and decl., and were similarly
binned into a $0.1\times0.2$ mag Hess diagram. We subtracted this from
the original cluster Hess map to define the cluster population.

We degraded the resulting photometry of each cluster using the
artificial stars generated for the \n4565 ACS field as described in
Section~\ref{sec:photometry}. For each artificial star that was
injected into the ACS image and run through the pipeline, both the
input and the recovered colors and magnitudes were recorded. We binned
these input magnitudes and colors into the same $0.1\times0.2$ mag
Hess diagram used for the clusters. To generate the degraded
photometry, we then randomly drew the same number of artificial stars
from each bin as were found in the equivalent bin of the final cluster
Hess map. If the selected artificial star was not successfully
recovered by the pipeline, then the cluster star was similarly
dropped. The measured photometry of the selected artificial stars was
then used for the final photometry. These data were combined for all
clusters in the same age bin.

Figure~\ref{fig:lmc} plots examples of these combined CMDs for the
youngest (left panel) and oldest (right panel) age bins, together with
the CMD of stars in the stellar-overdensity region of the \n4565 warp
(middle panel). In these CMDs, we shade both the AGB region and a
region that encompasses the young stars (MS, lHeB and uHeB). Evident
in even the young clusters ($\sim5$ Myr) are stars falling in the AGB
region. As discussed earlier, interstellar reddening in the \n4565
warp could further shift some of the stars identified as AGB stars in
the middle panel out of the shaded AGB region. However, this effect is
small ($<0.1$ mag) and will have a minimal effect on the number counts
($<10$\%).

\begin{figure*}
\begin{center} 
\resizebox{128mm}{!}{\includegraphics{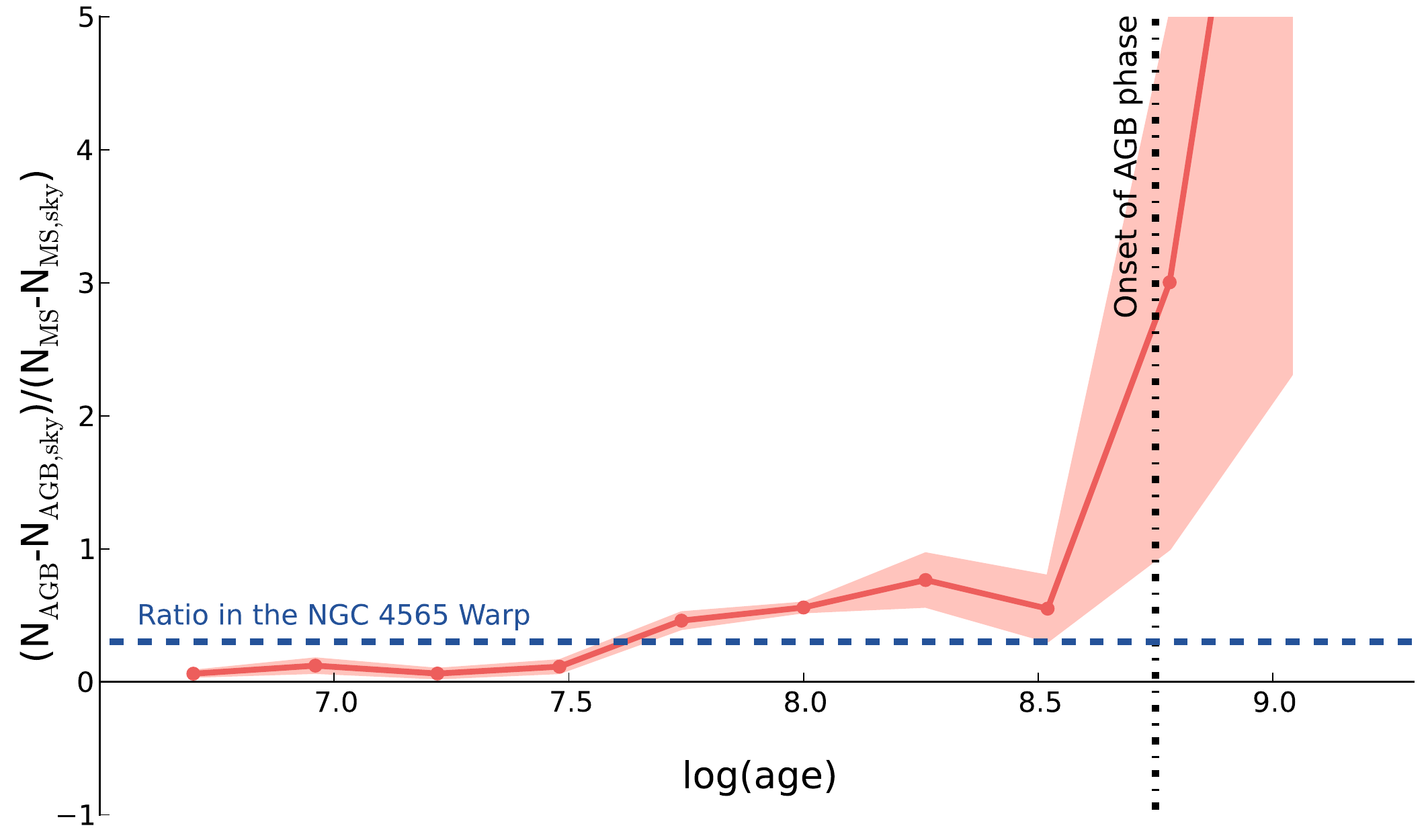}}
\caption{Ratio of the residual number of AGB stars to young stars from
  the LMC cluster analysis. This ratio is systematically greater than
  zero at all times and is consistent with the ratio seen in the
  \n4565 warp, indicated by a dashed blue line. The onset of AGB
  formation, shown as a vertical dot-dashed line, coincides with a
  dramatic increase in the ratio. The shaded region indicates the
  one-sigma spread of 100 Monte Carlo re-simulations of each age bin.}
\label{fig:ratio}
\end{center}
\end{figure*}

To assess the significance of the stars in the AGB region, we plot in
Figure~\ref{fig:ratio} the ratio of these stars to those in the
young-star region. To calculate the uncertainty on these values, we
run 100 Monte Carlo realizations of these data with random selections
from each artificial star bin. The 1$\sigma$ spread of these
realizations around the median values is indicated by a shaded region
in Figure~\ref{fig:ratio}. We note, as expected, a sharp increase in
this ratio around the onset of the AGB phase ($\sim 550$
Myr). However, even at younger ages the ratio is systematically
greater than zero and is consistent with the ratio found in the warp
of \n4565 ($\sim0.3$). Note that negative values for the ratio are
possible, as clusters where the density of a given population is
greater in the sky annulus than in the cluster are carried through as
negative detections. Hence, the systematic offset is likely real and
indicates that a spatial overdensity of young stars may result in a
slight overdensity of stars falling in the AGB region of the CMD that
likely are not AGB stars. Thus, the peak in the AGB profile of
Figure~\ref{fig:cross} may not necessarily be attributed to real AGB
stars but rather to a small number of younger stars.

\section{Implications for Warp\\Formation Mechanisms}
\label{sec:mechanisms}
Early work by \cite{hun69} showed that galaxy disks are unable to
sustain warps. Hence, the long-lived warps we see today are due to
ongoing gravitational effects. Initially, the dark matter halo was
believed to be the source of this excitation. By creating a rigid,
flattened halo that was misaligned with the disk, many authors were
able to create such warps
\citep[e.g.,][]{dek83,too83,spa88}.\footnote{Although, as noted by
  \cite{spa88}, such models cannot easily explain the turnover of the
  \n4565 warp at larger radii back to the disk plane.} However, such
halos are unrealistic, and more modern simulations with live halos
have shown that the induced warps quickly disappear, primarily due to
the shape of the inner halo realigning itself with the disk
\citep{dub95,bin98}.

Consequently, warps are now suspected to be driven by two broad
mechanisms: the tidal interaction of nearby galaxies and the effects
of misaligned infall.

\subsection{Tidal Interactions}
Interactions with satellite galaxies have long been suspected to
induce warps \citep[e.g.,][]{hua97} and have been investigated in
detail to explain the warp of the Milky Way (MW). However, the masses
of the Magellanic Clouds, specifically the LMC, seem insufficient to
generate the warp \citep{hun69}. \cite{wei98} proposed that the
response of the halo to the LMC may amplify the torque; this was
further modeled and found to be viable by \cite{tsu02} and
\cite{wei06}. However, despite this effect, the orbit of the LMC may
be at odds with the location of the existing warp. An analytical and
$N$-body simulation by \citep{gar00} suggests that the Magellanic
Clouds would induce a warp perpendicular to the orientation of the
actual warp. Instead, the favorable location of the Sagittarius dwarf
galaxy, which has been disrupted in our inner halo, may be the cause
of the MW warp \citep{iba98,bai03,gom13}.

Similar arguments for tidal satellite interactions causing warps have
been made for external galaxies such as \n5907, which was once thought
to be isolated \citep{sha98}. Additionally, gravitational encounters
with dark-matter substructure within the host halo have also been
shown to induce warps \citep{kaz08}. Hence, the two nearby companions
of \n4565, as seen in Figure~\ref{fig:ngc4565}, may in part be
responsible for the warp. Indeed, bending of the \hi{} layer toward
the closer IC 3571 system has been observed \citep{van05}. However,
this is independent of the larger warp.

If such tidal interactions are warping a pre-existing gas disk, then
they will also have an effect on any stars associated with that
disk. If such mechanisms formed the prodigious \hi{} warp of \n4565,
then they need to explain the lack of old stars associated with the
warp. The vertical energy of these dynamically hotter stars
will likely increase in response to a tidal interaction. However,
whether these stars can be sustained in a warped component, as opposed
to simply thickening the older stellar disk, remains to be shown by
simulations.

SF could also have been triggered by the same process forming the
warp. In such a chaotic event, resonances may selectively reinforce
the warp perturbation such that both the gas and the newly formed
stars spatially coincide, as seen in \n4565. If this is the case,
these tidal models may need to show that the warp in gas and stars can
survive and remain coincident for $\sim$1 Gyr, as potentially
indicated by the presence of the AGB stars described in
Section~\ref{sec:agb}.

\subsection{Misaligned Infall}
If simulations reveal the satellites of \n4565 are unable to induce
the warp of the host system, then misaligned infall remains the most
likely cause.

In a $\Lambda$CDM cosmology, material is regularly accreted from all
directions, causing the angular momentum of the host dark-matter halo
to continually, and significantly, change over time in a random walk
\citep{vit02}. \cite{ost89} proposed that the continual slewing of the
orientation of a flattened outer halo due to this misaligned accretion
causes warps in the galaxy disk. Indeed, such warping has been shown
in numerical simulations \citep[e.g.,][]{jia99,she06}. More generally,
\cite{deb99} showed that when the angular momentum of even a spherical
halo and galaxy disk differ, dynamical friction can cause long-lived
warps.

As with the tidal interactions, these simulations torque an existing
gas disk. Hence, in order to explain the stellar content of the \n4565
\hi{} warp, the relative influence of the formation mechanisms on
stars older than 1 Gyr needs to be further investigated in these
simulations.

Alternately, rather than torquing an existing disk, the warp itself
may have formed from the direct accretion of misaligned
material. \cite{san08} show several examples of gas-rich dwarf
galaxies interacting with their host galaxies, e.g., through an \hi{}
bridge. If these systems merge, with the higher-angular-momentum
material falling into an outer disk, we might expect a gas warp with
coincident stars. However, given that approximately 50\% of the
stellar mass in these dwarf galaxies is in place by $z=2$
\citep{wei11}, older stars would likely lie in the warp as well.

Instead of a dwarf galaxy, \cite{ros11} suggest that the warp forms
from infalling gas. In their models, this material cools and sinks
through a hot gaseous halo. The spin of this halo, which is not
aligned with the disk, torques the infalling material. Hence, the
angular momentum of the fresh gas when it finally reaches the disk is
aligned with the spin of the gas halo and thus forms a warped
disk. These hydrodynamic torques overwhelm the tidal torques from the
dark matter halo, which are insufficient for affecting the infalling
gas.

\cite{ros11} posit that the cooling gas in the long-lived warp will
reach densities sufficient for SF, resulting in young metal-poor stars
associated with the warp. Consequently, their Figure~14 shows a
remarkable agreement with that of \n4565 (Figure~\ref{fig:cmd}),
namely an old stellar component smoothly distributed around the
galaxy and a young stellar component associated with the warp.

Unlike the previous warp formation mechanisms, the \cite{ros11} warp
forms from pristine material. Hence, future metallicity measurements of
the warp may lend further credence to this model.

\section{Conclusions}
\label{sec:summary}
Using an $HST$ ACS observation, we have investigated the stellar
distribution of stars in the vicinity of the prodigious \hi{} warp of
\n4565. We find no correlation of old ($>$1 Gyr) stars with the
warp. Instead, these older stars lie in a symmetrical distribution
around the disk plane, likely due to a thick disk or flattened inner
halo. We do, however, find a significant population of young stars
($<$600 Myr) in the warp. An analysis of the SFH shows that the rate
of SF in the warp increased, relative to the surrounding regions, at
least 300 Myr ago. The sustained SFR over this time is
$(6.3^{+2.5}_{-1.5})\times10^{-5}$ \msol{} yr$^{-1}$ kpc$^{-2}$.

A slight excess of stars falling in the AGB region of the CMD are also
found in the warp. Stellar evolutionary models limit the age of these
stars to $<1$ Gyr, therefore placing an upper limit on the age of the
current warp. However, analysis of the CMDs of clusters in the LMC
suggests that these stars may not be true AGB stars but instead
younger stars with anomalous colors. If true, this may lower the age
of the warp to the order of hundreds of Myr.

The measured SFR implies a depletion time for the \hi{} gas in the
warp of $\sim60\pm20$ Gyr, similar to the rates found in the \hi{}
outer disks of nearby spiral galaxies. However, these outer disks are
often dominated by RGB populations
\citep[e.g.,][]{dejong07,rad12}. The lack of these stars in the \n4565
warp may suggest that the gas has been recently accreted. Indeed,
simulations of warp formation from newly accreted gas by \cite{ros11}
predict stellar distributions in qualitative agreement with the \n4565
\hi{} warp. In this scenario, the warp represents a large reservoir of
newly accreted \hi{} gas, which could be a potential fuel source for
the galaxy disk it circumscribes \citep[e.g., see discussion
in][]{san08,bigiel10a}

These observations help place constraints on numerical and analytical
models of warp formation. Future observations of the metallicity of
the warp will help clarify the origin of the gas and therefore the
mechanisms driving the warp.

\acknowledgments

The authors thank C. Purcell and V. Debattista for insightful
discussions on the theories of warp formation. Support for this work
was provided by NASA through grant GO-12196 from the Space Telescope
Science Institute, which is operated by the Association of
Universities for Research in Astronomy, Incorporated, under NASA
contract NAS5-26555.

{\it Facility:} \facility{HST (ACS)}

\end{document}